# Remote creation of strong and coherent emissions in air with two-color ultrafast laser pulses

Jinping Yao[1], Guihua Li[1,3], Chenrui Jing[1,3], Bin Zeng[1], Wei Chu[1], Jielei Ni[1,3], Haisu Zhang[1,3], Hongqiang Xie[1,3], Chaojin Zhang[1], Helong Li[2], Huailiang Xu[2,†], See Leang Chin[4], Ya Cheng[1,*], and Zhizhan Xu[1,#]

[1]*State Key Laboratory of High Field Laser Physics, Shanghai Institute of Optics and Fine Mechanics, Chinese Academy of Sciences, Shanghai 201800, China*

[2]*State Key Laboratory on Integrated Optoelectronics, College of Electronic Science and Engineering, Jilin University, Changchun 130012, China*

3*Graduate School of Chinese Academy of Sciences, Beijing 100080, China*

[4]*Center for Optics, Photonics and Laser (COPL) & Department of Physics, Engineering Physics and Optics, Université Laval, Quebec City, Qc G1V 0A6, Canada*

†*Corresponding author: huailiang@jlu.edu.cn*

\**Corresponding author: ya.cheng@siom.ac.cn*

#*Corresponding author: zzxu@mail.shcnc.ac.cn*

Abstract:

We experimentally demonstrate generation of strong narrow-bandwidth emissions with excellent coherent properties at ~391 nm and ~428 nm from $N_2^+$ ($B^2\Sigma_u^+$ ($v'=0$) $\rightarrow$ $X^2\Sigma_g^+$ ($v=0,1$)) inside a femtosecond filament in air by an orthogonally polarized two-color driver field (i. e., 800 nm laser pulse and its second harmonic). The durations of the coherent emissions at 391 nm and 428 nm are measured to be ~2.4 ps and ~7.8 ps respectively, both of which are much longer than the duration of the pump and its second harmonic pulses. Furthermore, the measured temporal decay characteristics of the excited molecular systems suggest an "instantaneous" population inversion mechanism that may be achieved in molecular nitrogen ions at an ultrafast time scale comparable to the 800 nm pump pulse.

**PACS** numbers: 42. 65. Re

Recently, lasing actions created remotely in air have attracted increasing interest due to their promising application in remote detection of multiple pollutants based on nonlinear spectroscopy [1-10]. Early experiments demonstrated remote ASE (amplified spontaneous emission) based lasers which have enabled operation either at ~391 nm and 337 nm using molecular nitrogen [3-5] or at ~845 nm using molecular oxygen [6] as gain medium. The generation of population inversion was ascribed to the recombination of free electrons with molecular nitrogen ions ($N_2^+$) [3-5] and resonant two-photon excitation of atomic oxygen fragments [6]. For the backward 845-nm ASE from atomic oxygen and the 337-nm ASE laser from neutral molecular nitrogen, the population inversion mechanisms are well understood [3-5, 11]. However, the mechanism responsible for the 391-nm ASE from $N_2^+$ is not totally clear, that is, the question as to how the population inversion in the ASE of the 391 nm is established is still open [4].

Remarkably, a series of recent experiments showed that strong and coherent multi-wavelength emissions with perfectly linear polarization (i. e., different from the random polarization of ASE) could be realized in nitrogen ($N_2^+$) and carbon dioxide ($CO_2^+$) gases using a wavelength-tunable OPA laser system with the wavelengths in the range of 1.2-2.4 μm, which can produce 3$^{rd}$ and 5$^{th}$ harmonics in air with spectral ranges overlapping the fluorescence lines of $N_2^+$ and $CO_2^+$ [7-9]. These emissions in $N_2^+$ (330, 357, 391, 428, 471 nm) and $CO_2^+$ (315, 326, 337, 351 nm) are found to be generated in an unexpected femtosecond timescale comparable to that of the pump lasers, indicating that population inversion in $N_2^+$ and $CO_2^+$ could have been achieved solely with intense ultrafast driver pulses. This observation challenges the previous conjecture on the population inversion mechanism based on the recombination of free electron with the molecular ions because such a process occurs on a timescale of a few nanoseconds [6]. To shed more light on the mechanisms underlying the ultrafast population inversion as well as on the coherent emissions themselves, which both are now under hot debate, temporal characterizations of these phenomena based on the concept of pump-probe measurement are important.

The fact that the ultrafast coherent emissions observed in previous experiments employing mid-infrared driver pulses always show a linear polarization parallel to that of the harmonic or supercontinuum indicates that a seeding effect may exist [7-9]. However, with the mid-infrared pump pulses, it is difficult to separate the self-generated harmonics or supercontinua from the driver pulses, making it difficult to vary the delay between the driver pulses and the seeding pulses. In this Letter, we will address this problem by remote generation of the strong and coherent emissions in air with an orthogonally polarized two-color laser field. In this new scheme, the driver pulses are provided by a 40 fs, 800 nm laser amplifier, whereas the 400 nm seed pulses are externally produced by a second harmonic generation process with a nonlinear crystal.

The pump-probe experiment scheme is illustrated in Fig. 1. A commercial Ti:sapphire laser system (Legend Elite-Duo, Coherent, Inc.), operated at a repetition rate of 1 kHz, provides ~40 fs (FWHM) laser pulses with a central wavelength at ~800 nm and a single pulse energy of ~6 mJ. The laser beam is firstly split into two arms using a 1:1 beam splitter with a variable delay: one is used as the pump beam (Pulse 1) and the other will pass through a 0.2-mm-thickness BBO crystal to produce the second harmonic probe pulse at 400 nm wavelength (Pulse 2) whose polarization is perpendicular to that of the pump pulses. The pump pulses have a pulse energy of ~1.9 mJ and a diameter of ~11 mm, whereas the probe pulses have a pulse energy of ~3 μJ and a diameter of ~6 mm, which are much weaker than the pump pulses. We have confirmed that the narrow-bandwidth emissions at 391 nm and 428 nm cannot be generated with the probe pulses alone. The pump and probe pulses are combined using a dichroic mirror with high reflectivity at 400 nm and high transmission at 800 nm, and then are collinearly focused by an *f = 40cm* lens into a chamber filled with 180 mbar of nitrogen gas to generate a filament and strong coherent emission. A small portion of the 800 nm beam split from the output beam of the laser system with an energy of 440 μJ (indicated as Pulse 3 in Fig. 1) is used for performing a cross-correlation measurement of the coherent emissions generated from the gas chamber. After passing through the gas cell, the 400 nm probe pulses containing strong coherent emissions are combined with Pulse 3 by another dichroic mirror (DM), and then are launched into a 2-mm-thick BBO crystal. The sum frequency generation (SFG) signal of the 800 nm and the coherent emission is produced and recorded by a grating spectrometer (Shamrock 303i, Andor) with a 1200 grooves/mm grating. The time-resolved SFG signal provides temporal information of the coherent emissions generated in $N_2$.

Figures 2(a) and (b) show two typical spectra measured in the forward propagation direction with the strong emissions generated respectively at the wavelengths of ~391 nm and ~428 nm in $N_2$. The emissions at the ~391 nm and ~428 nm correspond respectively to the transitions (0, 0) and (0, 1) between the vibrational levels of the excited state $B^2\Sigma_u^+$ and ground state $X^2\Sigma_g^+$ of $N_2^+$, as indicated in the inset of Fig. 1. In these two measurements, the BBO crystal for generating the second harmonic 400 nm laser light was finely tuned to optimize the 391 nm or 428 nm emissions, and the temporal and spatial overlap between the 800 nm and 400 nm pulses are optimized by maximizing the intensities of the strong emissions. It is also confirmed that when either the 800 nm pump beam or the 400 nm probe beam is blocked, the strong line emissions will disappear, indicating that both the pump and probe pulses are important for their creation. Furthermore, by placing a Glan-Taylor polarizer in front of the spectrometer, we examine the polarization of the strong line emissions at the ~391 nm and ~428 nm wavelengths. As indicated in the insets of Figs. 2(a) and (b), when the transmitted polarization direction is parallel to that of the 400 nm pulse, which is defined as 0 degree, both the ~391 nm and the ~428 nm emission are the strongest. On the contrary, when the polarizer is rotated by ±90 degrees, the emissions become too weak to be detected. Therefore, the line emissions at ~391 nm and ~428 nm are confirmed to have a nearly perfect linear polarization parallel to that of the second harmonic probe pulses. This important fact indicates that the weak second harmonic pulses play a role as a seed to activate the strong coherence emissions.

To gain a deeper insight, we investigate the intensities of the coherent emissions at both ~391 nm and ~428 nm as functions of the time delay between the pump and the probe pulses ($\tau_1$), as shown in Figs. 3(a) and (b), respectively. Here, the zero time delay is indicated by the green arrows in both Figs. 3(a) and (b) and the positive delay means that the second harmonic 400 nm probe pulse is behind the fundamental 800 nm pump pulse. As shown in Fig. 3(a), the emission at ~391 nm firstly increases rapidly on the timescale of ~400 fs (see inset of Fig. 3(a)), which reflects the long pulse duration of the second harmonic (~700 fs, see later), and then shows a slow exponential decay with a decay constant $\tau \approx 46.2 \, \text{ps}$, as indicated by the red dashed line. It is noteworthy that when the time delay is above ~1 ps, the pump pulses at 800 nm and the second harmonic probe pulses are essentially temporally separated, because the pulse durations of both the pump and probe pulses are significantly shorter than ~1 ps. However, even when the pump and probe pulses are temporally separated, the line emission at ~391nm can still be generated with perfectly linear polarization parallel to the 400 nm probe light. Not surprisingly, as most strong field molecular phenomena which are sensitive to molecular alignment and revival, we observe in this pump-probe experiment the modulation of the line emission at the times of $1/2 T_{rot}$, $1 T_{rot}$ and $3/2 T_{rot}$ ($T_{rot,}$ revival period of nitrogen molecules) [12, 13] as indicated in the inset of Fig. 3 (a). The mechanism behind this might be due to the modulation of the intensity of the probe pulses owing to the periodic focusing and defocusing in the filament due to the dynamic change of the alignment degree of the $N_2$ molecules [14, 15]. Figure 3(b) shows a similar decay behaviour of the line emission at ~428 nm, but with a much shorter decay time of ~2 ps.

Lastly, by introducing the third laser beam at 800 nm (Pulse 3), a cross-correlation measurement is performed to obtain the temporal information of the coherent line emissions at both ~391 nm and ~428 nm. Figures 4(a)-(c) show the frequency- and time-resolved SFG signals of the 800 nm and 400 nm probe pulses at ~267 nm, the 800 nm and the ~391 nm line emission at ~263 nm, and the 800 nm and the ~428 nm line emission at ~279 nm, respectively. We confirm that the narrow-bandwidth signals at ~263nm and ~279nm are unambiguously from the SFG of coherent line emissions and 800nm pulses based on the following two points. First, in comparison with the SFG signal of the 800 nm and the 400 nm probe pulses as shown in Fig. 4(a), both the SFG signal of 800 nm and the coherent emission at ~391 nm and the SFG signal of 800 nm and the coherent emission at ~428 nm, as shown in Figs. 4 (b) and (c) respectively, have much narrower spectra, because the coherent emissions of ~391 nm and ~428 nm have narrower bandwidths than the second harmonic 400 nm pulses. Second, the SFG signals at ~263nm and ~279nm cannot be observed in vacuum or argon. Here，the zero point of the time delay $\tau_2$ is defined as the point at which 800 nm and 400 nm probe pulses is well overlapped and the positive delay indicates that the second harmonic 400 nm probe pulse is behind the fundamental 800 nm pump pulse. It should also be pointed out that in order to obtain the three above-mentioned SFG signals, we have carefully adjusted the phase-matched angle, $\phi$, of the nonlinear crystal to optimize each SFG signal. Figure 4(d) presents the SFG signals distributed on the black dashed lines in Figs. 4(a)-(c) (i.e., 267.2 nm, 263.1 nm and 278.9 nm). It can be seen in Fig. 4 (d) that the SFG signals centered at 263.1 nm and 278.9 nm, which reflect the temporal profiles of the line emissions at 391 nm and 428 nm, start to rise gradually after the SFG signal centered at 267.2 nm (i. e., the contribution from the broad bandwidth 400 nm probe pulses and the 800 nm pulse). From the SFG signal centered at 267.2 nm, the pulse duration of 400 nm (FWHM) at the crystal is obtained to be ~ 700 fs due to the positive chirp induced by the dispersion in the windows, crystals *etc.*, and the cross phase modulation during filamentation. In contrast, the pulse durations of coherent emissions at ~391 nm and ~ 428 nm (FWHM) are ~2.4 ps and ~7.8 ps, respectively, which are much longer than that of the 400 nm probe pulses.

The mechanism responsible for the strong and coherent forward emissions is still to be clarified. Noticing that the polarization of the line emissions is determined by the polarization of the 400 nm probe pulses despite of their completely different pulse durations, a possible scheme of the seed amplification that can be enabled by generation of population inversion in $N_2^+$ is considered. In this situation, the population inversion has to be established within an ultrashort time period for initiating the amplification of the second harmonics, which are resonant with the transitions of electronic states in $N_2^+$. However, it is known that the ejection of an inner-valence electron (HOMO-2) of $N_2$ leaves the ion $N_2^+$ in the excited $B^2\Sigma_u^+$ state, whereas the ionization of an outer-valence electron (HOMO) leads to $N_2^+$ lying on the ground $X^2\Sigma_g^+$ state [16]. Although it has been observed experimentally that the lower-lying orbitals such as HOMO-1, HOMO-2 etc. indeed can participate in the ionization process [17, 18], numerical calculations [19-22] have shown that the ionization probability of HOMO-2 is about one to two orders of magnitude lower than that of HOMO in an intense laser field of the similar parameters as our experiment. Thus, there must be some other mechanisms for achieving the population inversion between the upper and lower levels if the seed-amplification scheme works. Because of the high laser intensity inside the filament, a nonlinear absorption process in $N_2^+$ ions in the ground state, as shown in Fig. 4(e), could occur, which induces the absorption of a few photons to deplete the population of $N_2^+$ in the lower vibrational levels of the ground state, and enhances the upper level of the B state with a Raman-type scheme, thus achieving the population inversion between B(0)-X(0) and B(0)-X(1).

With this population inversion scheme, the faster decay of the 428 nm emission than that of the 391 nm emission shown in Fig. 3 can be well understood. The vibrational relaxations, as indicated by the shortest green arrows in Fig. 4(e), first lead to an increase of the population on X(1) and then that on X(0) [23]. Thus, the cascade vibrational relaxation process makes the lifetime of the population inversion of B(0)-X(1) significantly shorter than that of B(0)-X(0), giving rise to the faster decay observed in Fig. 3(b) than that in Fig. 3(a).

In conclusion, we have observed strong and coherent emissions at ~391 nm and ~428 nm from nitrogen in an orthogonally-polarized two-color laser field, and measured their temporal profiles with cross-correlation measurements. We find that the pulse durations of the line emissions at both ~391 nm and ~428 nm are much longer than the 400 nm seed pulse, which is mainly due to the narrow bandwidths of the two line emissions. The results suggest that the coherent line emissions could originate from seed-injected amplification enabled by the remotely generated population inverted molecular systems in air.

This work is supported by the National Basic Research Program of China (Grant No. 2011CB808100), National Natural Science Foundation of China (Grant Nos. 11134010, 60921004, 11074098, 11204332 and 60825406), New Century Excellent Talent of China (NCET-09-0429), Basic Research Program of Jilin University, Canada Research Chairs, NSERC, DRDC Valcartier, CIPI, CFI, Femtotech and FQRNT.

Captions of figures:

Fig. 1 (Color online) Schematic of experimental setup. Inset: Energy-level diagram of $N_2$ and $N_2^+$ in which the transitions between $B^2\Sigma_u^+$ and $X^2\Sigma_g^+$ states are indicated with corresponding wavelengths.

Fig. 2 (Color online) Typical forward emission spectra with the coherent emission at (a) ~391 nm and (b) ~428 nm. Polarization property of coherent emissions at ~ 391 nm (Inset in (a)) and ~428 nm (Inset in (b)).

Fig. 3 (Color online) The strong coherent emission at (a) ~391 nm and (b)~428 nm as a function of time delay of the 800 nm pump and the 400 nm probe pulses. The zero delay is indicated by green arrows. Inset in (a): A higher resolution picture in the range from -1 ps to 13 ps.

Fig. 4 (Color online) Frequency- and time-resolved SFG signals of (a) 800 nm and the 400 nm probe pulse, (b) 800 nm and the coherent emission at ~391 nm and (c) 800 nm and the coherent emission at ~428 nm. (d) Time-resolved SFG signals distributed on black dashed lines in Figs. (a)-(c). (e) Schematic of pumping mechanism for generating population inversion.

Fig. 1

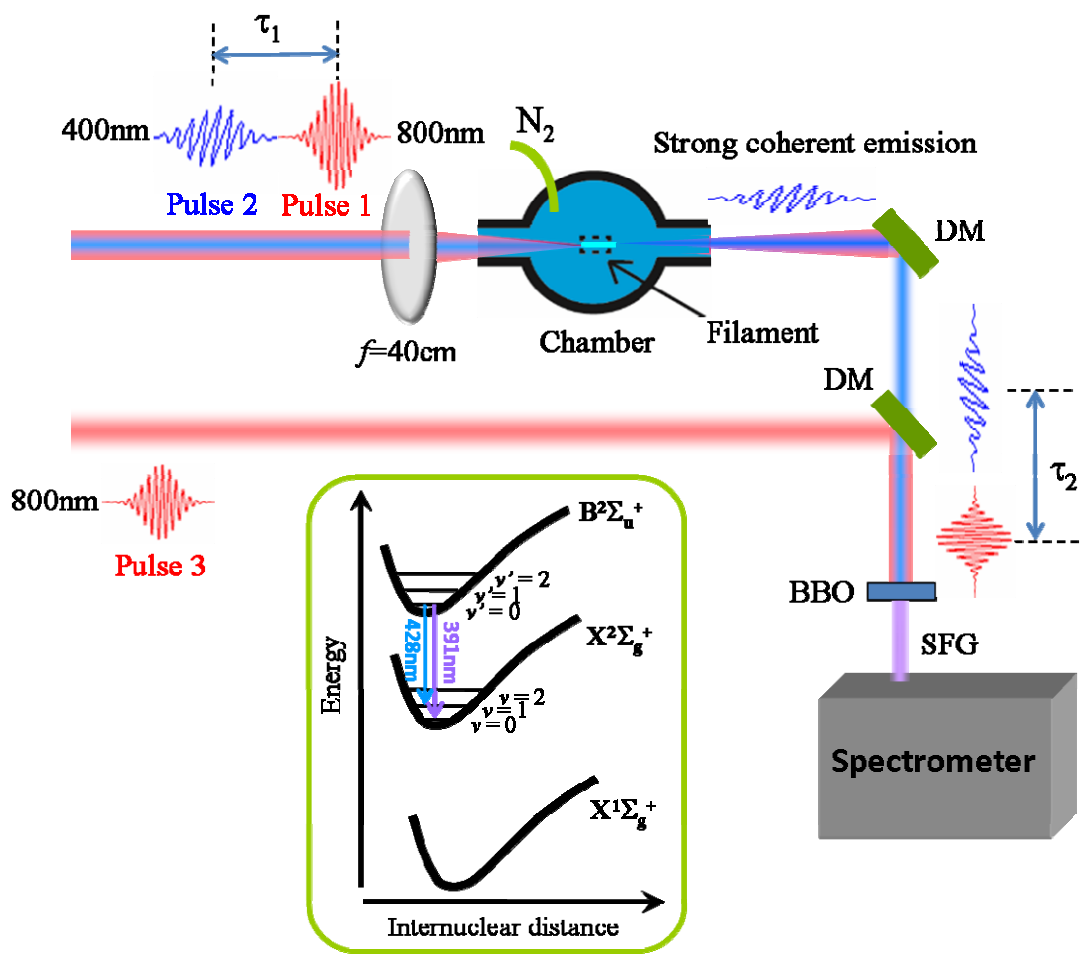

Fig. 2

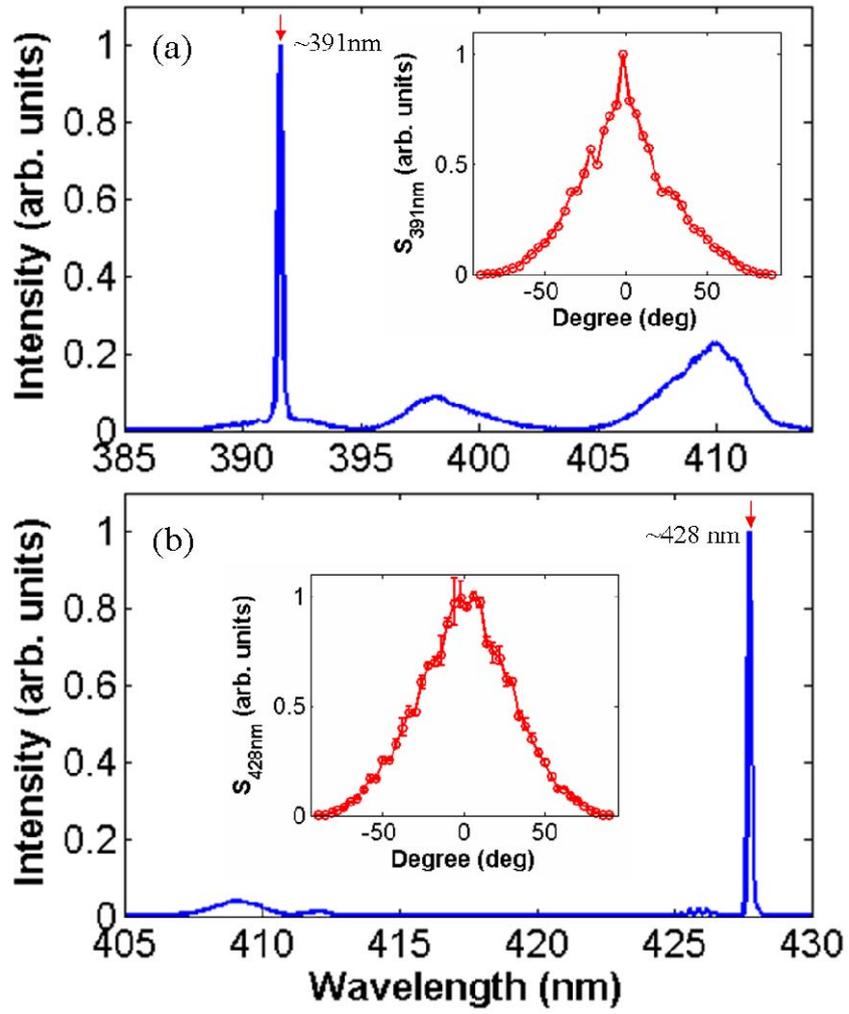

Fig. 3

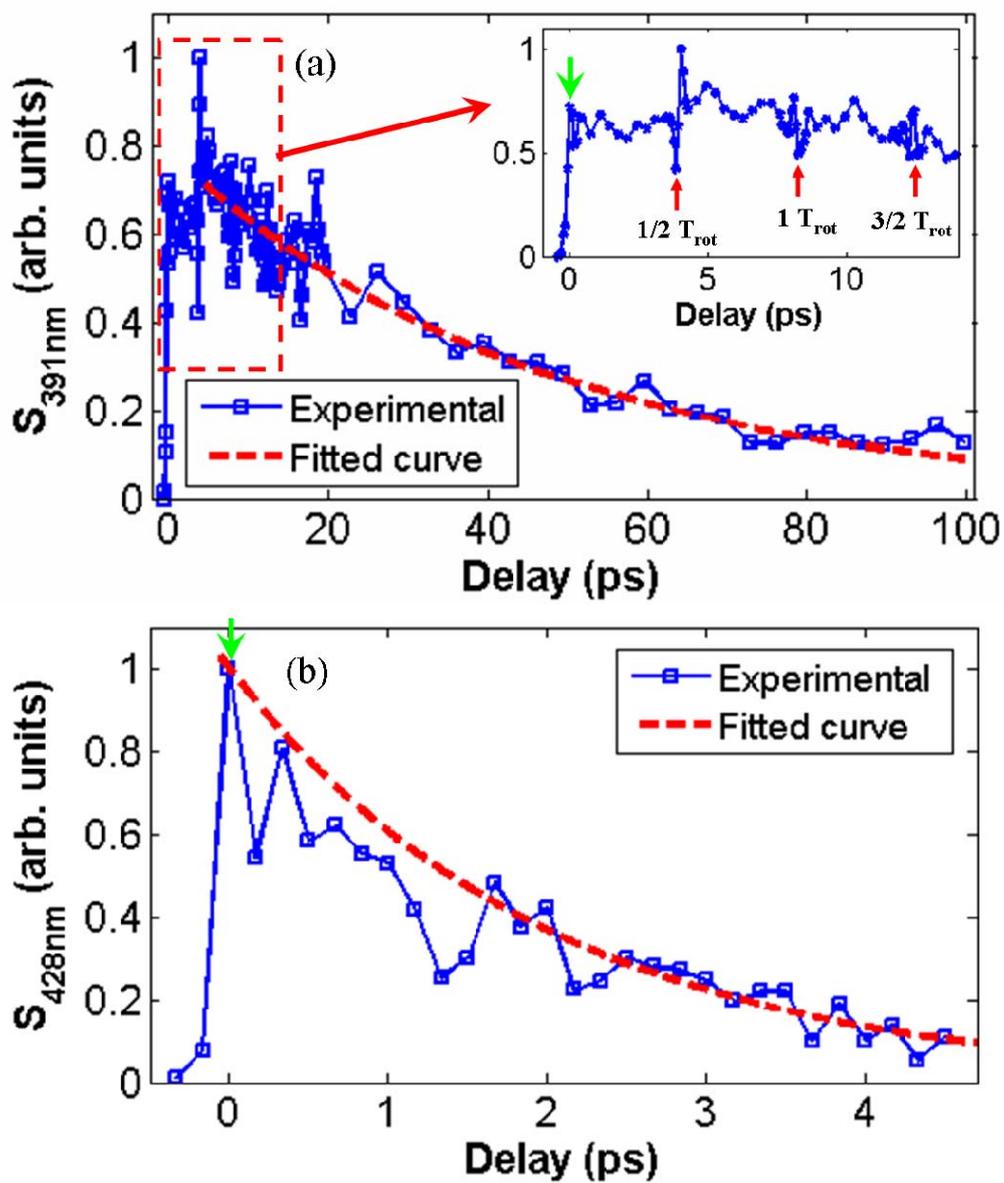

Fig. 4

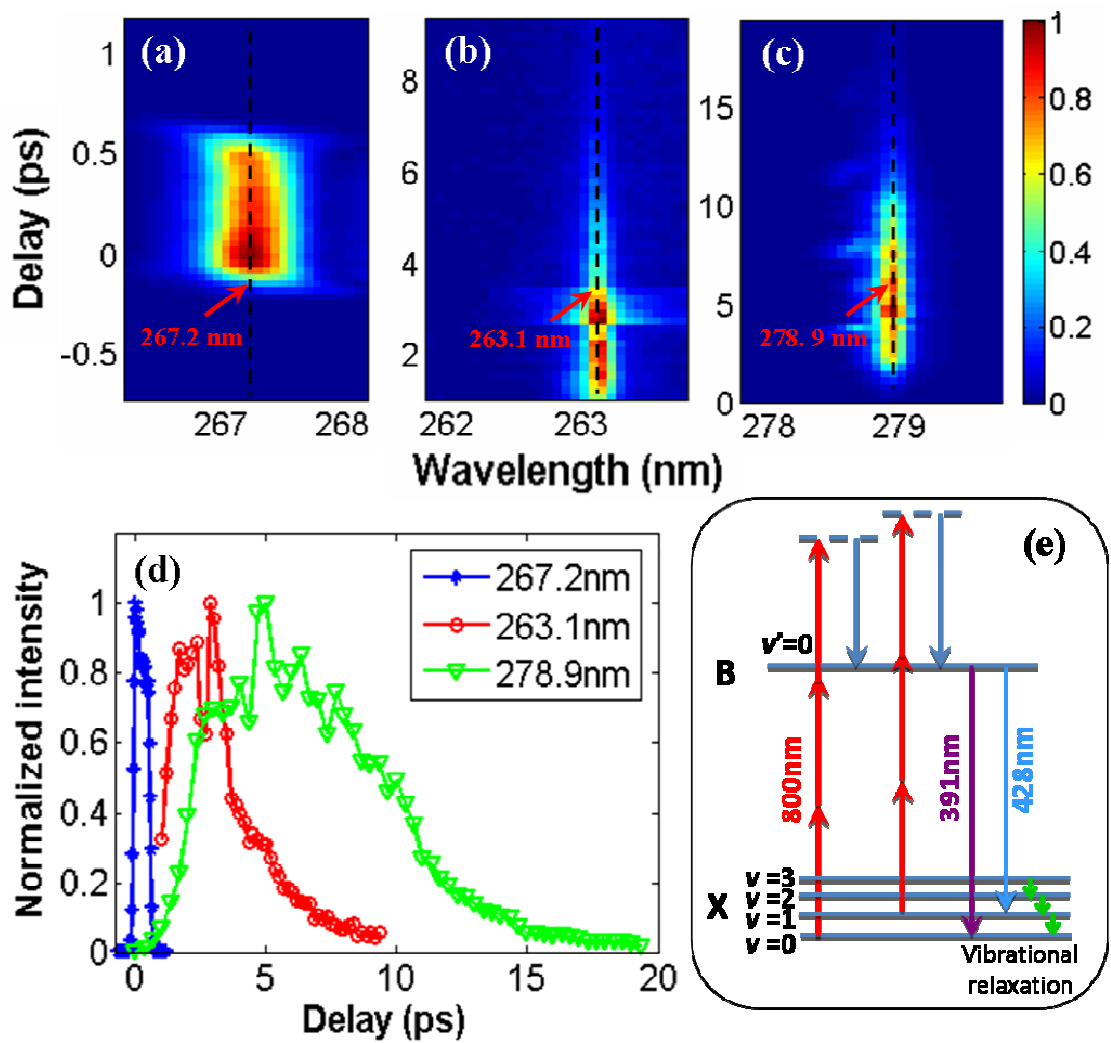